\documentstyle[aas2pp4]{article}

\slugcomment{Accepted for ApJ Letters Mar 25 1999}

\lefthead{Leggett, Toomey, Geballe, Brown}
\righthead{Revised Fluxes for Gliese 229B}
 
\begin{document}
 
\title{Revised Fluxes for Gliese 229B}

\author{S.K. Leggett}
\affil{Joint Astronomy Centre,  660 N. A'ohoku Place, Hilo, HI 96720
\nl skl@jach.hawaii.edu}
 
\author{D.W. Toomey} 
\affil{NASA IRTF, P.O. Box 4729, Hilo, HI 96720
\nl  toomey@irtf.ifa.hawaii.edu  }

\author{T.R. Geballe}
\affil{Gemini North Observatory, 670 N. A'ohoku Place, Hilo, HI 96720
\nl tgeballe@gemini.edu }

\and
 \author{R.H. Brown}
\affil{ Univ. of Arizona Lunar \& Planetary Lab., Tucson AZ 85721-0092
\nl rhb@lpl.arizona.edu}

\vskip 0.3truein
\centerline{Accepted March 25th 1999 for ApJ Letters}

\begin{abstract}
We have used the coronographic instrument CoCo with the IRTF's facility camera 
NSFCAM to obtain improved photometry at JHKL$^{\prime}$ of the giant 
planet/brown dwarf Gliese 229B.  We have recalibrated the
published spectra for this object, and re--calculated its luminosity.  Our 
L$^{\prime}$ value, and our flux calibration of the spectra at JHK, are 
significantly different from those previously published.  Our results show good 
agreement at all bands except H with evolutionary models by Burrows et al. 
which include grain condensation. The model comparison implies that Gliese 229B 
is likely to be a 0.5~Gyr--old 25 M$_J$ object with T$_{eff}$ $\sim$900~K.
\end{abstract}

\singlespace 
\section{Introduction}

\cite{nak95} discovered Gliese 229B during a coronographic search for low--mass 
companions to young M--dwarfs.  Despite continued searches, Gliese 229B to this 
date is the only substellar object that is resolvable and cool enough to show 
planet--like absorption features due to methane.  A good understanding of this 
object is key to our understanding of brown dwarfs and extra--solar planets in 
general.  Published work on this unique object have been primarily 
spectroscopic in nature (e.g. \cite{geb96,opp98}) or theoretical (e.g. 
\cite{all96,mar96,tsu96,bur97,gri98}).  The only published photometry for this 
object is 
that by \cite{mat96}, hereafter MNKO.  Photometry is vital for determining 
Gliese 229B's energy distribution (and hence luminosity and radius), as well as 
for flux calibrating the observed spectra.

In this paper we present new JHKL$^{\prime}$ photometry for Gliese 229B which 
is more accurate than the photometry obtained by MNKO.  We also re--flux 
calibrate the published spectra for the object and show that previous  flux 
calibrations are in error.

\section{Observations}

We obtained JHKL$^{\prime}$ photometry for Gliese 229B on 1999 January 14 using
the NASA Infrared Telescope Facility on Mauna Kea in Hawaii.  We used the 
facility infrared camera NSFCAM (\cite{shu94}) together with the coronographic
instrument CoCo (\cite{wan94,too98}).  CoCo is a cold coronographic attachment 
to NSFCAM that utilizes an apodized
focal plane mask and a pupil mask that stops the telescope outer diameter down 
to 80\% of the full aperture and  also obscures the inner 30\%.  The focal plane 
mask is about 2.5 arcseconds in diameter.  CoCo has an x/y adjustable pupil 
stop mask and a pupil imager for alignment.  Flux in the wings of a stellar 
profile are 
reduced by a factor of 5---10 when using CoCo.

We also made use of the new tip--tilt capability of the telescope.  The 
combination of the good image quality and tracking provided by the telescope 
with tip--tilt, and good removal of the light from the primary star Gliese 229A 
by CoCo, made our images much cleaner than those obtained by MNKO, with a 
resulting increase in photometric accuracy.

The observational technique used was to nod the telescope between Gliese 229A 
and a sky position 1 arcminute away.  The individual exposure time on 
Gliese 229B (and sky) at each of JHK was 90 seconds, and at L$^{\prime}$ it was
20 seconds.   The total on--target integration time was 6 minutes at J, 
7.5 minutes at H, 7.5 minutes at K and 20.7 minutes at L$^{\prime}$.  
Two JHK flux 
standards were also observed (UKIRT faint standards 4 and 12), and one
L$^{\prime}$ standard (HD 40335), with appropriate 
exposure times. The standards were placed on the same region of the detector as 
Gliese 229B, and were also observed by nodding to sky on alternate frames.  

The data were reduced by creating flatfields for every target and filter from 
the appropriate sky frames.  Individual flats had to be created as the mask 
could flex slightly at different telescope positions and hence change the 
overall ilumination.   At JHK each target frame was sky--subtracted and 
flatfielded.  Aperture photometry was carried out on each frame and the error 
taken to be the rms deviation in the set of Gliese 229B observations, combined 
with the scatter in the standard star values.  The combined error is 5\% at 
JHK. At L$^{\prime}$ we obtained two sets of data, separated by a standard star 
observation.  All the sky--subtracted and flatfielded frames in each set 
were combined (using sigma--clipping in IRAF).  The error was  
derived from the scatter in the remnant background levels, and is 10\%. 

Our JHK values agree with those of MNKO within the errors, but our  
L$^{\prime}$ value is three times brighter than their quoted value, which has 
very large errors.  Our value agrees well with model predictions (see \S 4).   
Table 1 gives our results, and Figure 1 shows our reduced images at J (a single 
90 second sky--subtracted and flatfielded image) and L$^{\prime}$ (the combined 
sky subtracted and flatfielded image for one of our two datasets, a total of 10 
minutes  on--target integration).  In Table 1 JHKL$^{\prime}$ are given on the
UKIRT system, which is identical to the IRTF system except at H such that
$$ (H - K)_{UKIRT} = 0.82[\pm 0.02] \times (H - K)_{IRTF} $$ (\cite{leg98}).

\begin{figure}[h!]
\figurenum{1}
\epsscale{50}
\plottwo{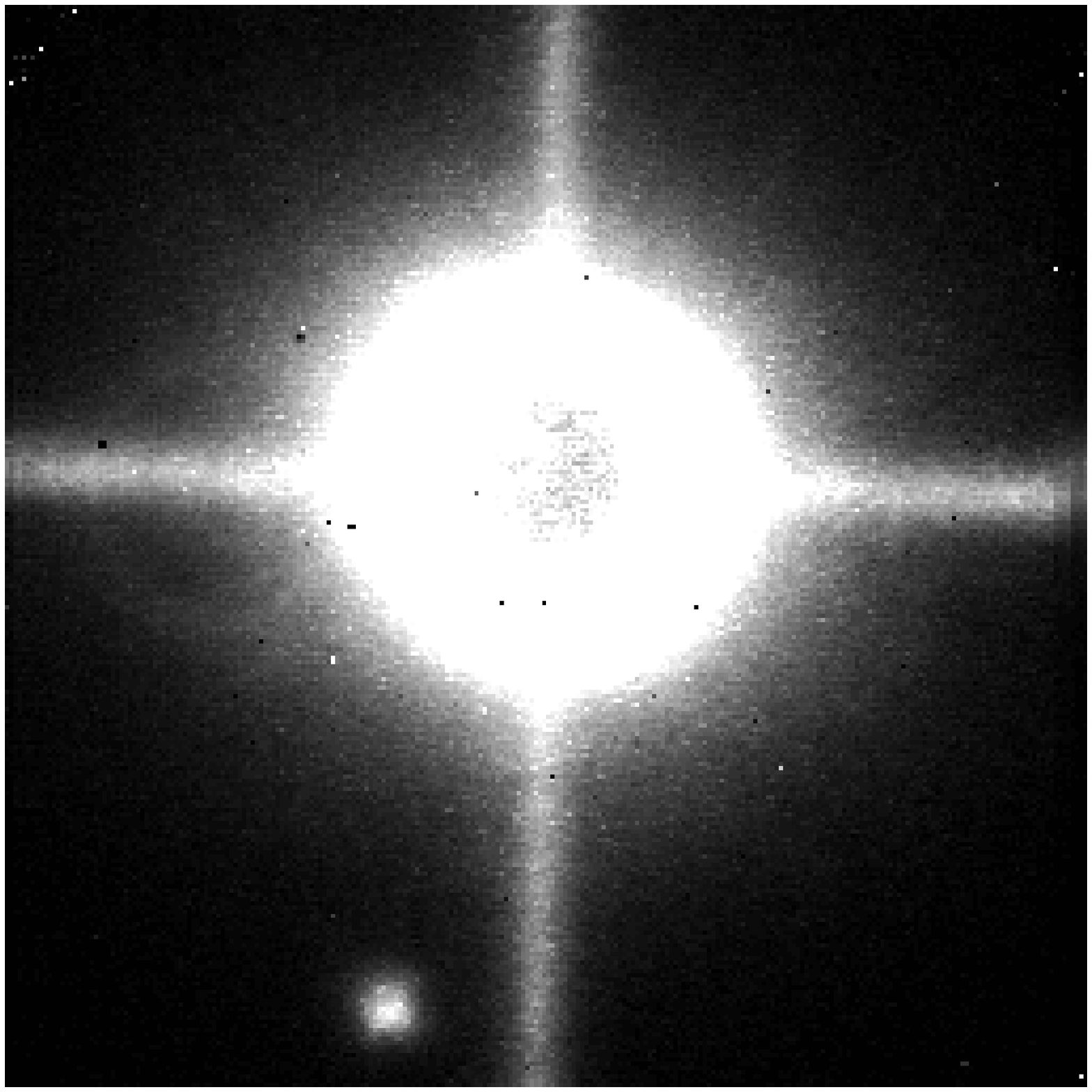}{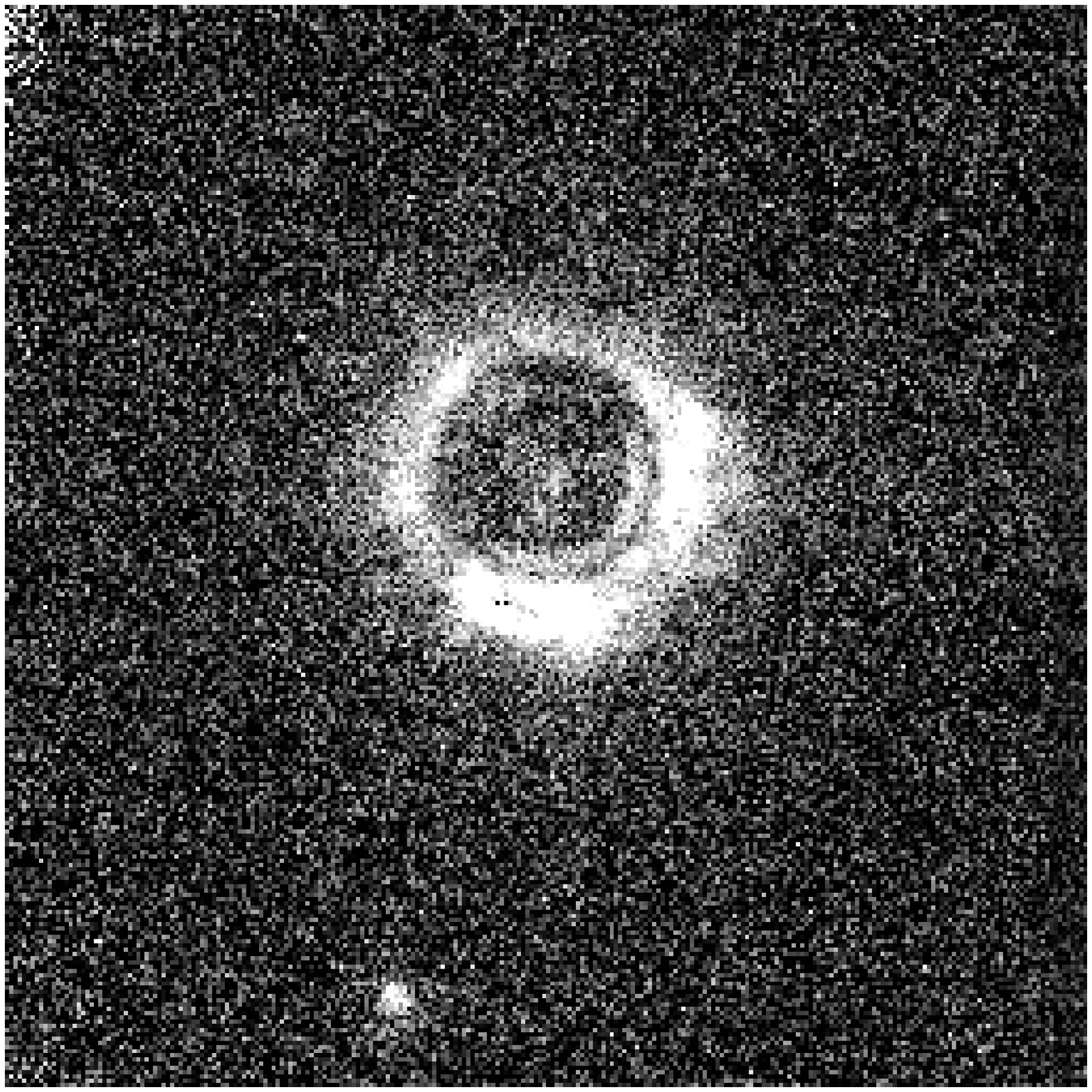}
\caption{NSFCAM and CoCo images of Gliese 229B at J (left) and  L$^{\prime}$ 
(right).  The image display stretch is zero to the Gliese 229B flux level.}
\end{figure}

\begin{deluxetable}{rrrrccr}
\tablecaption{PHOTOMETRY FOR GLIESE 229B \label{tbl-1}}
\tablehead{  
\colhead{J} & \colhead{H}  & \colhead{K} &  \colhead{L$^{\prime}$} & 
\colhead{M$-$m\tablenotemark{*}} & \colhead{Luminosity}& \colhead{BC$_K$} \nl 
\multicolumn{4}{c}{NSFCAM system} & \colhead{mag} & 
\colhead{log$_{10}$(L/L$_{\odot})$} & \colhead{mag} \nl 
} 
\startdata
14.32     & 14.35     & 14.42     & 12.18       & 1.185 &  $-$5.18 & 2.09 \nl
$\pm$0.05 & $\pm$0.05 & $\pm$0.05 & $\pm$0.10 & $\pm$0.070 &  $\pm$0.04 &  
$\pm$0.10 \nl
\tablenotetext{*}{\cite{van94}}
\enddata

\end{deluxetable}

\section{Flux Calibrated Spectra and Luminosity}

We have recalibrated the published spectra for Gliese 229B 
using our new photometry and the profiles of the NSFCAM filters.
The filter profiles are for the operating temperature of the camera,
and we have included the effect of the Mauna Kea atmospheric transmission.
 We have integrated the spectra over the JHKL$^{\prime}$ 
filters, and ratioed the flux to that obtained by integrating the observed Vega 
spectrum over the same filters.  Vega was adopted to be zero magnitude at all 
wavelengths,  and we scaled the Gliese 229B spectra such that the flux ratio 
through the filters matched our photometry.  The atmosphere--convolved
NSFCAM filter profiles, and the Vega spectrum, are available on request to 
S.~Leggett (skl@jach.hawaii.edu), for calibration of other spectra.

We found that although our JHK values are different by $<$10\% the published 
spectra in this region had to be multiplied by factors of 0.68 to 0.81,
and although the L$^{\prime}$ magnitude differed by a factor of three, the 
published spectra in 
that region only had to be multiplied by  a factor of 1.03.  The error in the 
original flux calibration of the \cite{opp98} spectra arose from their 
assumption that the target flux at the 
central wavelength of the filter could be simply scaled by the calibrator flux 
at that same wavelength.  This is not a valid assumption as the flux 
distribution across the filter is very different for Gliese 229B and any 
calibrator star.  In the case of the \cite{geb96} spectra, those authors did 
not attempt to scale by the observed photometry and their error arises from 
the usual problems of flux calibrating narrow slit
spectra combined with the uncertainty in removing the
scattered light from Gliese 229A.

Figure 2 shows the \cite{opp98} spectra for Gliese 229B.  We show the 
previously published spectral distribution, as well as our revised 
distribution. The NSFCAM filter profiles are overlaid to demonstrate the 
unusual flux weighting of the filters for Gliese 229B.  
Figure 3 shows the \cite{geb96} higher resolution 1---2.5$\mu$m spectrum, 
before and after recalibration. Scattered light from Gliese 229A was seen
in the raw data of both groups, despite their use of narrow slits. However
the lack of flux in the strong methane-- and water--absorption features
shows that any residual light from Gliese 229A has been accurately removed.

\begin{figure}[h!]
\figurenum{2}
\plotfiddle{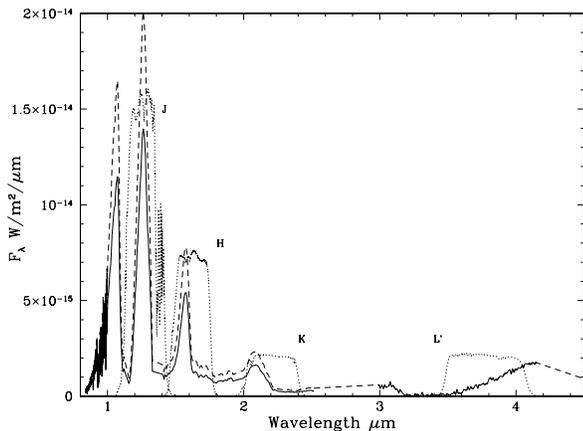}{1.7in}{-90}{30}{30}{-120}{160}
\caption{Gliese 229B spectrum from \cite{opp98} --- solid line shows the 
revised flux calibration, dashed as published.  The NSFCAM filter profiles are 
also shown.}
\end{figure}

\begin{figure}[h!]
\figurenum{3}
\plotfiddle{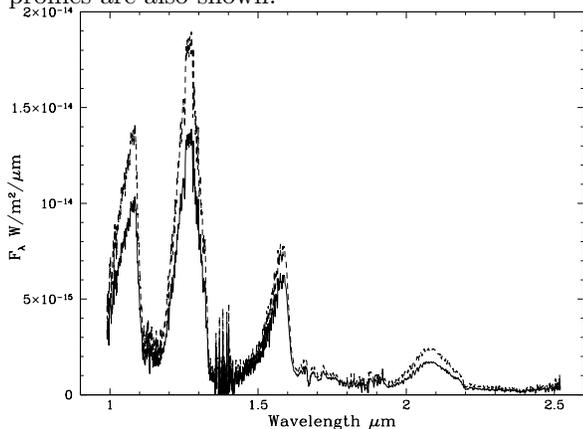}{1.7in}{-90}{30}{30}{-120}{160}
\caption{Gliese 229B spectrum from \cite{geb96} --- solid line shows the 
revised flux calibration, dashed as published.  }
\end{figure}

We have calculated the integrated luminosity using the recalibrated spectra.  
The unobserved region between the K--band and L--band was represented by a 
linear interpolation, which is consistent with the models. 
The L--band
spectrum ends at 4.15$\mu$m.  The integrated flux beyond this point
was adopted to be 29\% of the total integrated luminosity, as implied
by a recent model by Allard \& Hauschildt (private communication).  
The luminosity in this region is affected by strong CH$_4$ and H$_2$O bands 
at  6---8$\mu$m (unobservable from the ground), and this flux contribution is 
25\% smaller than that calculated by extending a  Rayleigh--Jeans tail from the
L spectrum.  The total flux is determined to be 6.39e-15 W/m$^2$ with a 
9\% error (due to the 
flux calibration errors and an estimated 20\% uncertainty in the flux beyond 
the L--band).  This flux gives the luminosity and bolometric correction
values listed in Table 1.  MNKO use models to estimate a bolometric correction
to their total in--band observed luminosity, and derive a total luminosity 
3\% smaller than ours (with an estimated 12\% error).

\section{Comparison to Models}

\cite{bur97} have published evolutionary models for giant planets and brown 
dwarfs.  This work includes the condensation of grains in the equation of state 
but does not include opacity due to grains, and it is to be expected that these 
models will be revised shortly.  Nevertheless our JHKL$^{\prime}$ colors and 
absolute magnitudes  show good agreement with these models in the color--color
diagrams (Burrows et al., Figures 20---26) at all bands except the 
H--band.   The discrepancy at H is large --- about 0.4 magnitudes.  However 
our three times brighter L$^{\prime}$ measurement ($\delta _{mag}=$ 1.2) now 
gives a result consistent with the J and K values, implying that Gliese 229B 
has a mass between 25 M$_J$ and 35 M$_J$ and an age between 0.5~Gyr and 1.0~Gyr.
The luminosity determined above and the Burrows et al. evolutionary tracks 
(their Figure 11) imply that a solution near the younger less--massive end of 
this range is more probable, and that the effective 
temperature is then around 900~K.  (MNKO also determined a temperature of 900~K
through a comparison of their photometry with a dust--free model by 
\cite{tsu96}.)

Burrows et al. suggest that the previous disagreement between the calculated 
and observed H and L$^{\prime}$ magnitudes were due to an incomplete CH$_4$ 
opacity  database.  However  CH$_4$ is an important opacity source at K
and L$^{\prime}$ both of which are now well matched by the models.  
Discrepancies are seen in the H--band with the hotter M-- and L--dwarfs also
(see e.g. \cite{leg98}) and these have been blamed on incomplete H$_2$O
opacity tables.  Presumably the problem is due to incomplete knowledge of
an opacity source, and the models, which are making large and rapid advances,
will soon be able to match the entire Gliese 229B spectrum.

\section{Conclusions}

We have obtained new JHKL$^{\prime}$ photometry using the coronographic 
instrument CoCo with NSFCAM on the IRTF.  The data are more accurate than 
previously published values, and while they agree within 10\%  at JHK, we have 
determined an L$^{\prime}$ magnitude three times brighter than that previously 
published.

We have re--flux--calibrated the published spectra for Gliese 229B by 
integrating the spectra over the NSFCAM filter profiles. We find that 
corrections of about 30\% are required at JHK.  The revised spectra are 
available in electronic form on request to B.~Oppenheimer 
(bro@astro.caltech.edu) and T.~Geballe (tgeballe@gemini.edu).

The JKL$^{\prime}$ colors and revised luminosity agree well with the 
evolutionary models of brown dwarfs by \cite{bur97}, implying that Gliese 229B 
is likely to be a 0.5~Gyr--old 25 M$_J$ object with T$_{eff}$ $\sim$900~K.  
We have not 
carried out a spectroscopic analysis as the models and synthetic spectra are 
currently being upgraded to include the effects of grains in both the equation 
of state and opacity tables --- and  grains have a large effect on the energy 
distribution of such cool objects.  We can look forward to a better 
understanding of the physics of atmospheres at these cool temperatures, now 
that there exists a more accurately flux calibrated spectrum for  Gliese 229B.

\acknowledgments
We are very grateful to the staff at IRTF.  The IRTF, 
the NASA Infrared Telescope Facility, is operated by the University of Hawaii 
under contract to NASA.   We are also grateful to France Allard and Peter 
Hauschildt for calculating the flux contribution longwards of 4.15$\mu$m, and
to the referee for useful comments.



\begin{thebibliography}{}
\bibitem[Allard et al. 1996]{all96} Allard, F., Hauschildt, P.H., Baraffe,  
   I. \& Chabrier, G. 1996, \apj, 465, L123
\bibitem[Burrows et al. 1997]{bur97}Burrows, A., Marley, M., Hubbard, W. B., 
Lunine, J. I., Guillot, T., Saumon, D., Freedman, R., Sudarsky, D. \& Sharp, C.
1997, \apj, 491, 856
\bibitem[Geballe et al. 1996]{geb96}Geballe, T.R., Kulkarni, S.R., Woodward, 
C.E., Sloan, G.C. 1996, \apj, 467, L101
\bibitem[Griffith, Yelle \& Marley 1998]{gri98} Griffith, C.A., Yelle, R.V. \& 
Marely, M.S. 1998, Science, 282, 2063
\bibitem[Leggett, Allard \& Hauschildt 1998]{leg98}Leggett, S.K., Allard, F. 
\& Hauschildt, P.H. 1998, \apj, 509, 836 
\bibitem[Matthews et al. 1996]{mat96}Matthews, K., Nakajima, T., Kulkarni,
 S.R., Oppenheimer, B.R., 1996, \aj,  112, 1678 (MNKO)
\bibitem[Marley et al. 1996]{mar96} Marley, M.S., Saumon, D., Guillot, T., 
Freedman, R.S., Hubbard, W.B., Burrows, A. \& Lunine, J.I. 1996, Science, 
272, 1919
\bibitem[Nakajima et al. 1995]{nak95} Nakajima, T., Oppenheimer, B.R., 
Kulkarni,
 S.R., Golimowski, D.A., Matthews, K. \& Durrance, S.T. 1995, \nat,  378, 463 
\bibitem[Oppenheimer et al. 1998]{opp98} Oppenheimer, B.R., Kulkarni, S.R., 
Matthews, K., Van Kerkwijk, M.H. 1998, \apj, 502, 932
\bibitem[Shure et al. 1994]{shu94}Shure, M.A, Toomey, D.W, Rayner, J.T.,
Onaka, P. \& Denault, A.J., 1994, Instrumentation in Astronomy VIII, (eds
Crawford, D.L \& Craine, E.R.), proc. SPIE 2198, 614
\bibitem[Toomey et al. 1998]{too98} Toomey, D.W., Ftaclas, C., Brown, R.H.
\& Trilling, D. 1998, proc. SPIE 3354, 782 
\bibitem[Tsuji et al. 1996]{tsu96}Tsuji, T., Ohnaka, K., Aoki, W., Nakajima, T.
1996, \aap, 308, L29
\bibitem[van Altena, Lee, \& Hoffleit 1994]{van94} van Altena, W.F., Lee, 
J.T., \& Hoffleit, E.D. 1994, The General Catalogue of Trigonometric Parallaxes 
(New Haven: Yale University Observatory)
\bibitem[Wang et al. 1994]{wan94} Wang, S., Owensby, P.D., Toomey, D.W.,
Brown, R.H., Stahlberger, W.E. \& Ftaclas, C. 1994,
proc. SPIE 2198, 578

\end{thebibliography}
\end{document}